\documentclass[floatfix,pra,twocolumn,showpacs,amsmath,amssymb,letterpaper]{revtex4}
\usepackage{times}
\usepackage{latexsym}
\usepackage{graphicx}
\usepackage{verbatim,times,bbm}
\usepackage{color}

\newcommand{\Tr}{{\mathrm{Tr}}}

\begin{document}

\title{Memory effects in attenuation and amplification quantum processes}
\author{Cosmo Lupo$^1$, Vittorio Giovannetti$^2$, and Stefano Mancini$^{1,3}$}
\affiliation{$^1$School of Science and Technology, University of Camerino, via Madonna delle Carceri 9, I-62032 Camerino, Italy \\
$^2$NEST, Scuola Normale Superiore and Istituto Nanoscienze-CNR, piazza dei Cavalieri 7, I-56126 Pisa, Italy\\
$^3$INFN-Sezione di Perugia, I-06123 Perugia, Italy}

\begin{abstract}

With increasing communication rates via quantum channels, memory
effects become unavoidable whenever the use rate of the channel is
comparable with the typical relaxation time of the channel
environment. We then introduce a model of bosonic memory channel,
describing correlated noise effects in quantum optical processes via
attenuating or amplifying media. To study such a channel model we
make use of a proper set of collective field variables, which allows
us to unravel the memory effects, mapping the $n$-fold concatenation
of the memory channel to a, unitarily equivalent, direct product of
$n$ single-mode bosonic channels. We hence estimate the channel
capacities by relying on known results for the memoryless setting.
Our findings show that the model is characterized by two different
regimes, in which the cross-correlations induced by the noise among
different channel uses are either exponentially enhanced or
exponentially reduced.

\end{abstract}


\maketitle

\section{Introduction}

Any quantum process can be described as a completely positive and
trace preserving map (quantum channel) acting on the set of trace
class operators (quantum states). In quantum information theory, one
of the most important problems is finding the maximal rates (i.e.,
the capacities) at which quantum or classical information can be
transmitted via quantum channels with vanishing error in the limit
of large number of signals (channel uses)~\cite{MH06}. Earlier works
on the subject focused on models where the noise affecting the
communication is assumed to act independently and identically for
each channel use (memoryless quantum channels). Recently, however,
an increasing attention has been devoted to correlated noise models
(quantum memory channels), see, e.g., \cite{KW2} and references
therein. Memory effects in the communication may arise when the
dynamics of the media is characterized by temporal correlations
which extend on timescales which are comparable with the time delay
between consecutive channel uses. For instance, under certain
conditions optical fibers may show relaxation times or birefringence
fluctuations times longer than the separation between successive
light pulses~\cite{exp}. Similar effects occur in solid state
implementations of quantum hardware, where memory effects  due to
low-frequency impurity noise produce substantial
dephasing~\cite{exp1}. Furthermore, moving from the model introduced
in~\cite{GiovMan}, memory noise effects have also been studied in
the contest of many-body quantum systems by relating their
properties  to the correlations of the channel environmental
state~\cite{PLENIO} or by studying the information flow in spin
networks~\cite{SPIN}.

It is generally believed that memory effects should improve the
information transfer of a communication line~\cite{Gall}. However,
finding optimal encodings is rather complex and up to date only a
limited number of models have been solved~\cite{DATTA1,KW2,VJP}. In
the context of bosonic channels \cite{ourprl}, we  have  recently
introduced an exactly solvable model which effectively describes the
transmission of  signals along attenuating media characterized by
finite relaxation times. It is characterized by two parameters $\mu,
\kappa\in [0,1]$, enabling to describe different scenarios ranging
from memoryless to inter-symbol interference channels~\cite{BDM}, up
to perfect memory configurations~\cite{bowen}, providing thus the
first comprehensive quantum information characterization of memory
effects in the continuous variable setup.

In this article we show the details of such a model and we extend it
to encompass amplification processes besides attenuation ones (this
is  formally obtained by extending the parameter space to include
also  $\kappa >1$ values). For such processes we estimate the
classical and the quantum capacity \cite{NOTA0} showing their
enhancement with respect to the memoryless case and proving that,
for the attenuating memory channel, coherent state encoding is
optimal for transmitting classical information. This is accomplished
by applying suitably chosen encoding and decoding unitary
transformations which allow us to \emph{unravel} the correlations
among different channel uses. In such a way the channel is mapped
into a unitary equivalent one, in which noises affecting different
channel uses are independent, although not identical \cite{trans}.
Using this mapping, the channel capacities can be estimated by
relying on known results on memoryless channels, which in the limit
of large number of channel uses provide converging lower and upper
bounds. Our results also allows us to point out the existence of two
different regimes  characterized by a threshold value of the product
$\mu\kappa$. Indeed, even in the presence of amplification, for
$\mu\kappa <1$  the correlations among different channels uses
induced by the channel dynamics are depressed exponentially with the
distance among the uses. This strongly resemble the behavior of the
so called {\em forgetful} memory channels, which were introduce in
Ref.~\cite{KW2} in the context of finite dimensional systems.
Vice-versa for $\mu\kappa > 1$ the memory effects, as well as the
correlations between different channel uses, are exponentially
enhanced by the amplification. Interestingly enough, irrespectively
of these rather different behaviors the capacities of the model show
no discontinuities at the critical threshold.

Even if the model we analyze is presented in an abstract setting, we
believe that the results presented in the paper, as typical in
optical communication analysis~\cite{CAVES}, may have a potential
impact on a variety of different realistic  setups which are
currently in use, ranging from fiber optics~\cite{GISIN}, to
wave-guide~\cite{TAN}, up to free-space signaling~\cite{ZEI}.
Moreover, the model can be applied for describing memory effects at
the interface between matter and light, e.g., in quantum repeaters
and quantum memories based on ensemble of atoms, in which the state
of the atomic ensemble is described in terms of an effective bosonic
degree of freedom \cite{interface}.

The article is organized as follows. In Sec.\ \ref{Model} we present
the memory channel model. In Sec.\ \ref{MUnraveling} we introduce
the technique of memory unraveling. In Sec.\ \ref{Asymptotics} we
discuss the asymptotic properties of the channel in the limit of
many channel uses and characterize the threshold. The memory
unraveling technique and the asymptotic properties of the channel
are hence used for evaluating the classical and quantum capacity in
Sec.\ \ref{Capacities}. In Sec.\ \ref{Forget} we argue on the
forgetfulness of the model. Finally, Sec.\ \ref{Conclusion} we
present concluding remarks.

\section{The model}\label{Model}

\begin{figure}
\centering
\includegraphics[width=0.35\textwidth]{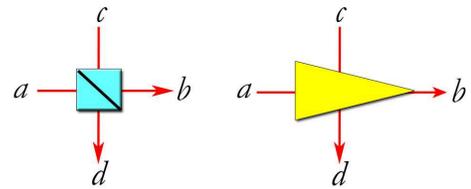}
\caption{(Color online.) Basic building blocks of the Gaussian
memory channel, mixing the signal (represented by the horizontal
lines) and the idler mode (vertical lines). On the left: the
circuital representation of the beam-splitter transformations in
Eq.s (\ref{BS}). On the right: the circuital representation of the
linear amplifier in Eq.s (\ref{LA}).} \label{bblocks}
\end{figure}

The most relevant effects for optical communications are the
attenuation and amplification processes, which are modeled as an
exchange interaction of each ingoing mode (signal) with a
corresponding external bosonic mode (idler), modeling a local
environment. We define a model of Gaussian memory channel by
concatenating these basic transformations. An attenuation process
involving a pair of ingoing modes, described through the ladder
operators $\{ a, a^\dag \}$ and $\{ c, c^\dag \}$, and the
corresponding outgoing modes $\{ b, b^\dag \}$, $\{ d, d^\dag\}$, is
described by a unitary transformation of the form
$U_\mathrm{BS}=\exp{(\theta a^\dag c - \mathrm{h.c.})}$, modeling a
beam-splitter mixing the two modes. In the following we assume the
parameter $\theta$ to be real and positive, yielding the
Heisenberg-picture transformations
\begin{subequations}\label{BS}
\begin{eqnarray}
b = {U_\mathrm{BS}}^{\hspace{-0.05cm}\dag} \, a \, U_\mathrm{BS} = \cos{\theta} \, a -\sin{\theta} \, c \, , \\
d = {U_\mathrm{BS}}^{\hspace{-0.05cm}\dag} \, c \, U_\mathrm{BS} =
\cos{\theta} \, c +\sin{\theta} \, a \, ,
\end{eqnarray}
\end{subequations}
together with their hermitian conjugates. Different choices for the
parameter $\theta$ do not affect the main results of our analysis.
The quantity $\cos^2{\theta}$ is usually referred to as the
beam-splitter transmissivity. Analogously, we will consider the
linear parametric amplifier, described by a unitary operator of the
form $U_\mathrm{LA} = \exp{(\chi a^\dag c^\dag - \mathrm{h.c.})}$.
For simplicity we assume the parameter $\chi$ to be real and
positive, even though other choices do not substantially alter our
analysis. In the Heisenberg picture such a transformation acts on
the ladder operators as follows
\begin{subequations}\label{LA}
\begin{eqnarray}
b = {U_\mathrm{LA}}^{\hspace{-0.1cm}\dag} \, a \, U_\mathrm{LA} = \cosh{\chi} \, a +\sinh{\chi} \, c^\dag \, , \\
d = {U_\mathrm{LA}}^{\hspace{-0.1cm}\dag} \, c \, U_\mathrm{LA} =
\cosh{\chi} \, c +\sinh{\chi} \, a^\dag \, ,
\end{eqnarray}
\end{subequations}
together with the hermitian conjugates. The quantity $\cosh^2{\chi}$
denotes the gain of the linear amplifier. The circuital
representations of these basic unitary transformations are in
Fig.~\ref{bblocks}.

To define the action of the quantum memory channel, we consider a
sequence of $n$ consecutive channel uses. Such a sequence is
associated with a collection of $n$ bosonic modes with ladder
operators $\{ a_j, {a_j}^{\hspace{-0.05cm}\dag} \}_{j=1,\dots n}$,
representing the channel inputs, and the corresponding channel
outputs associated with the ladder operators $\{ b_j,
{b_j}^{\hspace{-0.05cm}\dag} \}_{j=1,\dots n}$. We model the channel
environment as a collection of environmental modes with $\{ e_j,
{e_j}^{\hspace{-0.05cm}\dag} \}_{j=1,\dots n}$. If the use rate of
the channel becomes high enough to induce unwanted overlaps between
successive uses, or to interfere with the finite relaxation time of
the channel environment, then on a given use the local environment
could results still ``contaminated" by the previous uses. To account
for such effects, in the same spirit of the Ref.s~\cite{KW2,bowen},
we introduce a {\em memory} mode that connects all the local
environmental modes via an exchange interaction mechanism.

\begin{figure}
\centering
\includegraphics[width=0.35\textwidth]{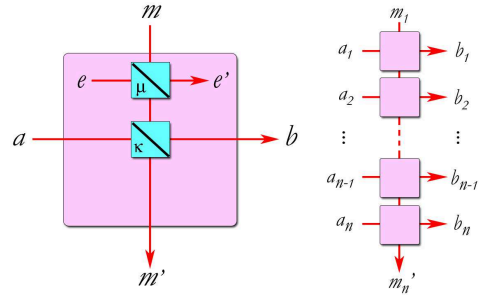}
\caption{(Color online.) Left: a single use of the attenuating
memory channel is described as an elementary transformation which is
the concatenation of two beam-splitters, respectively characterized
by the transmissivities $\mu$ and $\kappa$. The first beam-splitter
couples the memory mode with the local environment, the second one
mixes the memory mode with the input signal. Right: $n$ uses of the
memory channel are described as the $n$-fold concatenation of the
elementary transformation \cite{KW2,bowen}. The concatenation is
obtained by identifying, for any $j$, the outgoing memory mode at
the $j$-th channel use with ingoing memory mode at the $(j+1)$-th
channel use. The memoryless limit is obtained when there is no
information flow via the memory mode, i.e., for $\mu=0$.}
\label{lossym}
\end{figure}

\begin{figure}
\centering
\includegraphics[width=0.35\textwidth]{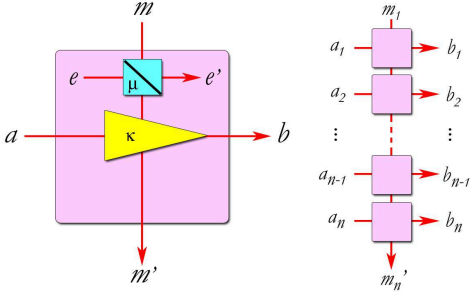}
\caption{(Color online.) Left: a single use of the amplifying memory
channel is described as an elementary transformation which is the
concatenation of a linear amplifier with gain $\kappa > 1$ and a
beam-splitter with transmissivity $\mu$. Right: the $n$-fold
concatenation of the amplifying memory channel is obtained by
identifying, for any $j$, the outgoing memory mode at the $j$-th
channel use with ingoing memory mode at the $(j+1)$-th channel use
\cite{KW2,bowen}.} \label{amplifym}
\end{figure}

A single use of the memory channel is modeled by a network of linear
elements: beam-splitters and/or linear amplifiers being the basic
building blocks. Such an elementary transformation is depicted in
Fig.\ \ref{lossym} for the lossy memory channel, and in Fig.\
\ref{amplifym} for the amplifier one \cite{network}. In the
Schroedinger picture it is formally described by a map
\begin{equation}\label{elementary}
\Phi\left(\rho_{a,m}\right) = \Tr_e \left[ U
\left(\rho_{a,m}\otimes\rho_e\right) U^\dag\right] \, ,
\end{equation}
where $\rho_{a,m}$ is the ingoing state of the input and memory
mode, and $\rho_e$ denotes the initial state of the local
environment. The unitary $U$ is the composition of two beam-splitter
unitaries for the attenuating channel (Fig.\ \ref{lossym}), or the
composition of one beam-splitter and one linear amplifier for the
amplifying memory channel (Fig.\ \ref{amplifym}). Finally, the
partial trace over the local environment accounts for the fact that
its final state is ignored. In the Heisenberg picture, this
elementary transformation transforms the ladder operators according
to $b=U^\dag a U$, $m'=U^\dag m U$. For the case of the attenuating
memory channel, the elementary transformation is the composition of
two beam-splitters with transmissivities $\mu$, $\kappa$
(transmissivities are within the interval $[0,1]$), hence we have
\begin{subequations}\label{lossy1}
\begin{align}
m' & = \sqrt{\mu\kappa} \, m + \sqrt{1-\kappa} \, a + \sqrt{(1-\mu)\kappa} \, e \, , \label{lossy1m}\\
b & = \sqrt{\kappa} \, a - \sqrt{(1-\mu)(1-\kappa)} \, e \, -
\sqrt{\mu(1-\kappa)} \, m \, ,
\end{align}
\end{subequations}
together with the hermitian conjugate relations. Analogously, the
elementary transformation in the attenuating memory channel is the
composition of a beam-splitters with transmissivities $\mu$, and a
linear amplifier with gain $\kappa$, with $\kappa \in (1,\infty)$.
The ladder operators hence transform according to
\begin{subequations}\label{amplify1}
\begin{align}
m' & = \sqrt{\mu\kappa} \, m + \sqrt{\kappa-1} \, a^{\dag} + \sqrt{(1-\mu)\kappa} \, e \, , \label{amplify1m}\\
b & = \sqrt{\kappa} \, a + \sqrt{(1-\mu)(\kappa-1)} \, e^\dag +
\sqrt{\mu(\kappa-1)} \, m^{\dag} \, ,
\end{align}
\end{subequations}
together with the hermitian conjugates.

The action of the memory channel upon $n$ uses is obtained by
identifying the outgoing memory mode at each channel use with the
ingoing memory mode at the following one (see Fig.s \ref{lossym},
\ref{amplifym}). Formally, this is accomplished by assigning a map
of the form
\begin{equation}\label{mapn}
\Phi_n[\rho^{(n)}_{a,m}] = \Tr_e \left\{ \mathcal{U}_n
\left[\rho^{(n)}_{a,m} \otimes \rho^{(n)}_e\right]
\mathcal{U}_n^\dag \right\} \, ,
\end{equation}
where $\rho^{(n)}_{a,m}$ denotes the ingoing state of the $n$
channel input modes and of the memory mode at first channel use, and
$\rho^{(n)}_e$ is the state of the $n$ local environments. The
unitary operator appearing in (\ref{mapn}) is $\mathcal{U}_n = U_n
U_{n-1} \cdots U_1$, with $U_j$ being the unitary describing the
elementary transformation in Eq.\ (\ref{elementary}) applied at the
$j$-th channel use. In the Heisenberg picture the outgoing operators
$m_n'$, $b_j$'s (for $j=1,\dots n$), together with their hermitian
conjugates, verify the identities $m'_n = \mathcal{U}_n^\dag m_1
\mathcal{U}_n$, $b_j = \mathcal{U}_j^\dag a_j \mathcal{U}_j$. For
the attenuating memory channel ($\kappa \leqslant 1$), by iterating
Eq.s~(\ref{lossy1}), we have
\begin{widetext}
\begin{eqnarray}
m'_n & = & \sqrt{\mu\kappa}^n \, m_1 +
\sqrt{1-\kappa}\,\sum_{j=1}^{n} \sqrt{\mu\kappa}^{n-j} \, a_{j} +
\sqrt{(1-\mu)\kappa}
\sum_{j=1}^{n} \sqrt{\mu\kappa}^{n-j} \, e_{j} \, , \label{m_lossy} \\
b_j & = & \sqrt{\kappa} \, a_j +
\sqrt{\mu}\,(\kappa-1)\sum_{h=1}^{j-1} \sqrt{\mu\kappa}^{j-h-1} \,
a_{h} - \sqrt{(1-\mu)(1-\kappa)}\,\sum_{h=1}^{j}
\sqrt{\mu\kappa}^{j-h} \, e_{h} -
\sqrt{\mu(1-\kappa)}\sqrt{\mu\kappa}^{j-1} \, m_1 \, .
\label{a_lossy}
\end{eqnarray}
Similarly, for the linear amplifier memory channel ($\kappa > 1$),
from Eq.s~(\ref{amplify1}) we get
\begin{eqnarray}
m'_n & = & \sqrt{\mu\kappa}^n \, m_1 + \sqrt{\kappa-1}\,\sum_{j=1}^n
\sqrt{\mu\kappa}^{n-j} \, a_j^\dag +
\sqrt{(1-\mu)\kappa}\,\sum_{j=1}^n \sqrt{\mu\kappa}^{n-j} \, e_j \, ,\label{m_ampl}\\
b_j & = & \sqrt{\kappa} \, a_j +
\sqrt{\mu}\,(\kappa-1)\sum_{h=1}^{j-1} \sqrt{\mu\kappa}^{j-h-1} \,
a_h + \sqrt{(1-\mu)(\kappa-1)}\,\sum_{h=1}^j \sqrt{\mu\kappa}^{j-h}
\, e_h^\dag + \sqrt{\mu(\kappa-1)}\,\sqrt{\mu\kappa}^{j-1} \,
m_1^\dag \, . \label{a_ampl}
\end{eqnarray}
\end{widetext}
The Heisenberg-picture relations (\ref{m_lossy})-(\ref{a_ampl}) will
be the starting point of the information-theoretical
characterization of the memory channel which will be the aim of the
following sections.

Differently from other models where the causal structure is not
manifest \cite{PLENIO,GiovMan}, this construction leads to a {\em
non-anticipatory} channel~\cite{Gall} where a given input can only
influence subsequent channel outputs (i.e.\ for each $j$, $b_j$
depends only upon the $a_{h}$'s with $h\leqslant j$). The
transmissivity $\mu$ clearly plays the role of a {\it memory
parameter}. It can be related to the ratio between the time delay
$\Delta t$ between to successive channel uses and the typical
relaxation time $\tau$ of the channel environment~\cite{VJP}: for
instance we may identify $\mu \simeq \exp{(-\Delta t/\tau)}$. In
particular, the model reduces to a memoryless (attenuating or
amplifying) channel~\cite{HolevoWerner} for $\mu=0$ (the input $a_j$
only influences the output $b_j$), and to a channel with perfect
memory \cite{bowen} for $\mu=1$ (all $a_j$'s interacts {\em only}
with the memory mode). These two limiting settings respectively
correspond to the regime $\Delta t \gg \tau$, and $\Delta t \ll
\tau$. Intermediate configurations are associated with values $\mu
\in (0,1)$ and correspond to {\it inter-symbol interference}
channels, for which the previous input states affect the action of
the channel on the current input~\cite{BDM}. Of particular interest
is also the case $\kappa=0$ where $\Phi_n$ describes a {\em quantum
shift} channel~\cite{BDM}, where each input state is replaced by the
previous one.

Finally, to exhaustively define the channel model, we have to fix
the initial state of the local environmental modes $\{ e_j, e_j^\dag
\}$. Different choices for the environment states lead to channels
with different features. For instance, an environment in a
correlated state leads to an additional source of correlated noise,
similarly to the effect described by the models considered in
\cite{GiovMan,lbmc,weakforget}. In the following we assume the local
environments to be in the vacuum state. With this choice, the noise
caused by the interaction with the local environmental modes is
limited to the shot-noise. Moreover, as we will show, this choice
for the environmental states allows us to {\it unravel} the
correlation in the memory channel \cite{trans}, and to compute
exactly the quantum and classical capacity by using known results
for the memoryless setting \cite{Wolf,broadband}.

\section{Unraveling the memory}\label{MUnraveling}

The goal of this article is to estimate the classical and quantum
capacity of the model of bosonic Gaussian memory channel presented
above. As a first step in this direction, we consider a block of $n$
successive channel uses and show that the application of suitably
defined encoding and decoding unitary transformations allows us to
{\it unravel} the memory effects, by mapping $n$ uses of the memory
channel into the direct product of $n$ uncorrelated channels acting
on a suitable set of collective variables \cite{trans}. Such a map
defines a decomposition into `normal modes' of the bosonic memory
channel which, under certain conditions, is a common feature of
bosonic systems described by quadratic Hamiltonians (see, e.g.,
\cite{KB}).

Moreover, we ought to distinguish among four different notions of
(classical and quantum) capacities of the memory channel, depending
whom the memory mode is assigned to~\cite{KW2}. Specifically, the
initial and final state of the memory mode can be under the control
of the sender, of the receiver, or can be ignored by both and
assigned to the environment. Here we only consider the latter case,
and procrastinate the discussion of this issue to Sec.\
\ref{Forget}.

As a first step, we notice that Eq.s (\ref{a_lossy}), (\ref{a_ampl})
can be written in the following compact form
\begin{subequations}\label{collective_1}
\begin{align}
b_j = \sum_{h} A_{jh} \, a_h - \sum_{h} E_{jh} \, e_h \, , \quad & (\kappa \leqslant 1) \, , \\
b_j = \sum_{h} A_{jh} \, a_h + \sum_{h} E_{jh} \, e_h^{\dag} \, ,
\quad & (\kappa > 1) \, ,
\end{align}
\end{subequations}
where we have defined $e_0 := m_1$, and we have introduced the
matrices $A$, $E$, whose elements are readily obtained from
(\ref{a_lossy}), (\ref{a_ampl}). In particular, the causal structure
of the memory channel implies $A_{jh}=E_{jh}=0$ for $j < h$. By
increasing values of $n$, two sequences of matrices of increasing
dimension are defined. For each $n$, we consider the {\it singular
value decomposition} of the matrix $A$, that is,
\begin{equation}
A_{jh} = \sum_{j'=1}^n O_{jj'} \, \sqrt{\eta^{(n)}_{j'}} \, O'_{j'h}
\, ,
\end{equation}
where $O$, $O'$ are unitary matrices of size $n$. Actually they can
be assumed to be real orthogonal, for, in our construction, the
matrix $A$ has real entries. It follows from Eq.s
(\ref{collective_1}) that the singular value decomposition of the
matrix $E$ reads
\begin{equation}
E_{jh} = \sum_{j'=1}^n O_{jj'} \, \sqrt{|\eta^{(n)}_{j'}-1|} \,
O''_{j'h} \, ,
\end{equation}
and that $\eta^{(n)}_j \leqslant 1$ for $\kappa \leqslant 1$, and
$\eta^{(n)}_j \geqslant 1$ for $\kappa \geqslant 1$.

We hence define the following set of collective output variables
\begin{equation}
\mathrm{b}_j := \sum_{j'=1}^n O_{j'j} \, b_{j'} \, ,
\end{equation}
and the input and environmental collective variables
\begin{subequations}
\begin{eqnarray}
\mathrm{a}_j &:=& \sum_{j'} O'_{jj'} \, a_{j'} \, , \label{a_collective}\\
\mathrm{e}_j &:=& \sum_{j'} O''_{jj'} \, e_{j'} \, .
\end{eqnarray}
\end{subequations}
These variables are named `collective' since they are delocalized
over different channel uses. By construction they satisfy the
canonical commutation relations
$[\mathrm{b}_j,{\mathrm{b}_{j'}}^{\hspace{-0.1cm}\dag}] =
[\mathrm{a}_j,{\mathrm{a}_{j'}}^{\hspace{-0.1cm}\dag}] =
[\mathrm{e}_j,{\mathrm{e}_{j'}}^{\hspace{-0.1cm}\dag}] =
\delta_{jj'}$. Moreover, it follows from Eq.s (\ref{collective_1})
that they verify the identities
\begin{subequations}\label{unraveled}
\begin{align}
\mathrm{b}_j  & = \sqrt{\eta_j^{(n)}} \,  \mathrm{a}_j -
\sqrt{1-\eta_j^{(n)}} \, \mathrm{e}_j \, , \quad (\kappa \leqslant 1) \, , \\
\mathrm{b}_j  & = \sqrt{\eta_j^{(n)}} \,  \mathrm{a}_j +
\sqrt{\eta_j^{(n)}-1} \, \mathrm{e}_j^\dag \, , \quad (\kappa > 1)
\, .
\end{align}
\end{subequations}

\begin{figure}
\centering
\includegraphics[width=0.45\textwidth]{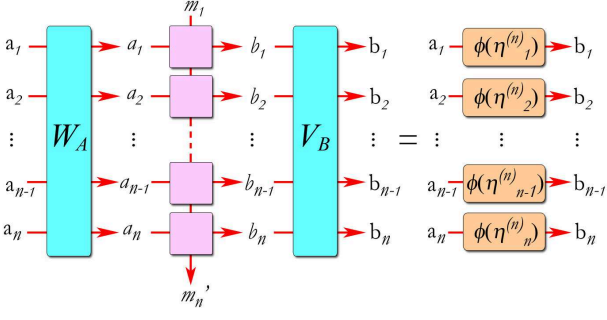}
\caption{(Color online.) The figure shows a circuital representation
of Eq.\ (\ref{equivalence}). Provided that the environmental modes
and the initial memory mode are in the vacuum state, $n$ uses of the
memory channel are unitary equivalent to the direct product of $n$
channels independently acting on the collective modes.}
\label{unraveling}
\end{figure}

We denote $W_A$, $V_B$, $T_E$ the canonical
unitaries~\cite{HOLEVOBOOK} that implement the transformations $a_j
\rightarrow \mathrm{a}_j = {W_A}^{\hspace{-0.1cm}\dag} \, a_j \,
W_A$, $b_j \rightarrow \mathrm{b}_j = {V_B}^{\hspace{-0.1cm}\dag} \,
b_j \, V_B$ and $e_j \rightarrow \mathrm{e}_j =
{T_E}^{\hspace{-0.05cm}\dag} e_j T_E$. We have hence shown that the
channel $\Phi_n$ is unitarily equivalent to the map
\begin{equation} \label{nuova}
\Phi^\prime_n[\rho_{a,m}^{(n)}]= \Tr_e \left\{
\mathcal{U}_{n}^{\prime} \left[ \rho_{a,m}^{(n)} \otimes
{\rho_e^{(n)}}^\prime \right]{\mathcal{U}_{n}^\prime}^\dag \right\}
\, ,
\end{equation}
with ${\rho_e^{(n)}}^\prime := {T_E}^{\hspace{-0.05cm}\dag} \,
\rho_e^{(n)} \, T_E$, and where the unitary transformation
$\mathcal{U}_{n}^\prime := V_B \mathcal{U}_{n} (W_A \otimes T_E)$
induces the linear transformations in Eq.s (\ref{unraveled}).
Formally, the unitary equivalence reads
\begin{equation}\label{equivalence}
\Phi^\prime_n[\rho_{a,m}^{(n)}] =  V_B \, \Phi_n[
{W_A}^{\hspace{-0.1cm}\dag} \, \rho_{a,m}^{(n)} \, W_A] \,
{V_B}^{\hspace{-0.1cm}\dag} \, ,
\end{equation}
that is, we can treat the output states of $\Phi_n$ as output of
$\Phi_n^\prime$ by first counter-rotating the input
$\rho_{a,m}^{(n)}$ by $W_A$ (coding transformation) and then by
rotating the output by $V_B$ (decoding)~\cite{GiovMan}. Assuming
then $\rho_e^{(n)}$ to be the vacuum state, we have
${\rho_e^{(n)}}^\prime = \rho_e^{(n)}$ and the map~(\ref{nuova}) can
be written as the direct product of independent bosonic channels,
i.e.,
\begin{equation}\label{product}
\Phi^\prime_n = \bigotimes_{j=1}^n \phi[\eta_j^{(n)}],
\end{equation}
with $\phi[\eta_j^{(n)}]$ being the single-mode transformations
whose Heisenberg-picture description is given in Eq.s
(\ref{unraveled}). A schematic representation of the encoding and
decoding procedure which unravels the memory channel is shown in
Fig.~\ref{unraveling}.

In conclusion, we have shown that the $n$-fold concatenation of the
memory channel is unitarily equivalent to: $n$ single-mode amplifier
channels, each with gain $\eta_j^{(n)} \geqslant 1$ for $\kappa
\geqslant 1$; $n$ single-mode attenuating channels, each with
transmissivity $\eta_j^{(n)} \leqslant 1$ for $\kappa \leqslant 1$;

\section{Asymptotical properties}\label{Asymptotics}

The aim of this section is to describe the sequence of parameters
$\{ \eta^{(n)}_j\}_{j=1,\cdots, n}$ in the limit of $n\to\infty$. The
regularity of the sequence may allow us to determine converging
upper and lower bounds on the channel capacities, whose evaluation
will be the subject of the following section.

For all $n$, the parameters $\eta^{(n)}_j$ are the $n$ eigenvalues
of the real, positive matrix
\begin{equation}
M^{(n)} := A A^\dag \, .
\end{equation}
From Eq.s (\ref{a_lossy}), (\ref{a_ampl}), we obtain the following
expression:
\begin{equation}\label{theseq}
M^{(n)}_{jj'} = \delta_{jj'} + \left(\kappa_{jj'}-1\right)
\sqrt{\mu\kappa}^{|j-j'|} \, ,
\end{equation}
where
\begin{equation}
\kappa_{jj'} := \kappa + \mu(\kappa-1)^2
\sum_{h=0}^{\min{\{j,j'\}}-2} (\mu\kappa)^h \, .
\end{equation}
The asymptotic behavior of the sequence of matrices $M^{(n)}$
strongly depends upon the value of the product $\mu\kappa$. Such
parameter quantifies the relation between the memory mode at two
successive channel uses (see Eq.s~(\ref{lossy1m}),
(\ref{amplify1m})), and allows us to split the parameter region in
distinct sectors: on one hand,  for $\mu\kappa<1$  the information
carried by the memory mode is attenuated and $M^{(n)}$ results in a
convergent sequence of bounded operators; on the other hand, for
$\mu\kappa\geqslant 1$ the influence of the memory mode is amplified
and the sequence does not converge. In the following we will refer
to these regions as {\it below threshold} ($\mu\kappa < 1$),  {\it
above threshold} ($\mu\kappa > 1$), and {\it at threshold}
($\mu\kappa =1$).

\subsection{Below Threshold}

For $\mu\kappa<1$ the sequence of matrices $M^{(n)}$ is {\it
asymptotically equivalent} \cite{toeplitz} to the (infinite)
Toeplitz matrix $M^{(\infty)}$, whose elements are
\begin{equation}
M_{jj'}^{(\infty)} := M_{j-j'}^{(\infty)} = \delta_{jj'} +
\left[{\kappa^{(\infty)}}-1\right] \sqrt{\mu\kappa}^{|j-j'|} \, ,
\end{equation}
with
\begin{equation}
\kappa^{(\infty)} := \lim_{\min{\{j,j'\}}\rightarrow \infty }
\kappa_{jj'} = \kappa + \frac{\mu (\kappa-1)^2}{1-\mu\kappa} \, .
\end{equation}
Following Ref.~\cite{toeplitz} the  asymptotic distribution of the
eigenvalues of the matrix $M^{(n)}$  can then  be expressed   in
terms of the continuous function obtained by Fourier transforming
the elements of the matrix $M^{(\infty)}$, i.e.,
\begin{equation}\label{mono_spectrum}
\eta(z) = \sum_{j=-\infty}^\infty M^{(\infty)}_{j} e^{iz j/2} = \left|
\frac{\sqrt{\mu}-\sqrt{\kappa}\,e^{iz/2}}{1-\sqrt{\mu\kappa}\,e^{iz/2}}
\right|^2 \, ,
\end{equation}
with $z\in[ 0, 2\pi]$~\cite{NOTA1}. The connection between the
parameters $\eta^{(n)}_j$ and the function (\ref{mono_spectrum}) is
formalized by the Szeg\"o theorem \cite{toeplitz} which states that,
for any smooth function $F$, we have
\begin{equation}\label{szego}
\lim_{n\to\infty} \frac{1}{n} \sum_{j=1}^n F[\eta^{(n)}_j] = \int_0^{2\pi}
\frac{dz}{2\pi} F[\eta(z)] \, .
\end{equation}

\subsection{Above Threshold}

When the channel operates above threshold the sequence of matrices
does not converge. For $\mu \kappa> 1$  we find it convenient to
rewrite Eq.~(\ref{theseq}) as the sum of two terms:
\begin{equation}\label{commuting}
M^{(n)} = c^{(n)} P^{(n)} + \Delta{M}^{(n)} \, ,
\end{equation}
where $P^{(n)}$ is a sequence of rank one projectors, $c^{(n)}$ is a
diverging sequence of positive real numbers, and $\Delta{M}^{(n)}$
is a sequence of matrices which asymptotically converges towards the
(infinite) Toeplitz matrix $\Delta{M}^{(\infty)}$, with entries
\begin{equation}\label{Mrescaled}
\Delta{M}^{(\infty)}_{jj'} = \frac{(1-\mu)(\kappa-1)}{\mu\kappa-1}
\frac{1}{\sqrt{ \mu\kappa }^{|j-j'|}} \, .
\end{equation}
The explicit expressions of the projectors $P^{(n)}$ and of the
matrices $\Delta{M}^{(n)}$ are reported in Appendix~\ref{APPE}. One
can easily verify that in the asymptotic limit of $n\rightarrow
\infty$ they commute (or, to say it more formally, that their
commutator is asymptotically equivalent to the null matrix). We
conclude that the spectrum of the matrices (\ref{theseq}) is
asymptotically composed of one diverging eigenvalue [from the
diverging sequence $c^{(n)}$] and of the asymptotic distribution of
the eigenvalues of the infinite Toeplitz matrix (\ref{Mrescaled}).
Similarly to the below threshold case, the latter can be calculated
by Fourier transforming: we hence find that it is described by the
function in Eq.~(\ref{mono_spectrum}), extended to the region
$\mu\kappa>1$.

\subsection{At threshold}

At the threshold value, $\mu\kappa=1$, the matrix $M^{(n)}$ can be
expressed as
\begin{equation}\label{theseqTHR}
M^{(n)}_{jj'} = \delta_{jj'} + (1-\mu) + \frac{(1-\mu)^2}{\mu} \min\{j,j'\} \;.
\end{equation}
A part from the trivial case $\mu=1$ where $M^{(n)}$ coincides with
the identity operator (perfect channel), the analysis of the
asymptotic behavior is rather cumbersome (for instance, it is not
possible to identify a single diverging eigenvalue). We thus resort
on numerical diagonalization of the sequence of matrices. This shows
that the finite part of the spectrum is again well fitted by the
distribution Eq.~(\ref{mono_spectrum}), extended to the manifold
with $\mu\kappa =1$.

\section{Capacities}\label{Capacities}

Equation~(\ref{product}) suggests the possibility of computing the
classical and quantum capacity of the memory channel by applying the
results for the memoryless multi-mode channels
\cite{broadband,Wolf}. To do so however, we have first to deal with
the fact that the single-mode channels forming $\Phi^\prime_n$ are
not necessarily identical. Therefore the map~(\ref{product}) is not
memoryless in a strict sense. To cope with this problem we will
construct two collections of memoryless channels which upper and
lower bound the capacity of the memory channel, and then we will use
the asymptotic distribution~(\ref{mono_spectrum}) to show that, for
large $n$, they converge toward the same quantity.

We proceed as follows. For $n$ uses of the memory channel we arrange
the single-mode channels $\{ \phi(\eta^{(n)}_j)\}$ in such a way
that the corresponding parameters $\eta^{(n)}_j$ are ordered
monotonically. We further divide the single-mode channels in $P$
blocks, each containing $\ell = n/P$ channels, and define the
infimum  and supremum for each block, i.e.\
\begin{align}
\underline{\eta}^{(P)}_p = & \inf_n \inf_{(p-1)\ell < j \leqslant
p\ell} \eta^{(n)}_j \, , \\
\overline{\eta}^{(P)}_p = & \sup_n \sup_{(p-1)\ell < j \leqslant
p\ell} \eta^{(n)}_j \, .
\end{align}

\begin{figure}
\centering
\includegraphics[width=0.35\textwidth]{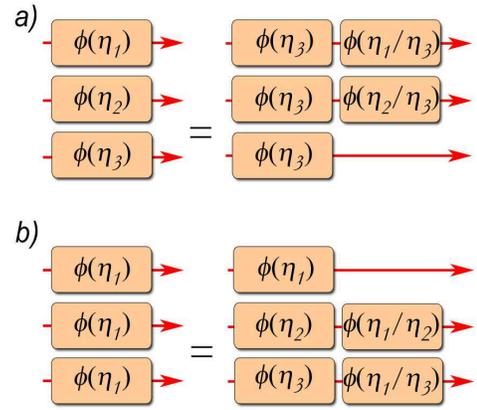}
\caption{(Color online.) The figure shows a circuital representation
of the law of composition of attenuating or amplifying bosonic
channels. To fix the ideas, let us consider a sequence of $n = 3$
uses of the memory channel, associated to three parameters $\eta_j$.
For the attenuating channel, let us assume $\eta_1 \leqslant \eta_2
\leqslant \eta_3 \leqslant 1$. It follows that: (a) the rate of
information transmission cannot be greater than the one of the
channel with all the transmissivities equal to $\eta_3$; and (b) it
cannot be smaller than the one with all the transmissivities equal
to $\eta_1$. The same bounds hold true for the amplifying channel,
assuming $\eta_1 \geqslant \eta_2 \geqslant \eta_3 \geqslant 1$.}
\label{bounds}
\end{figure}

It is worth noticing that, for any integer $P$ and independently on
$n$, the two collections of parameters, $\{ \overline{\eta}^{(P)}_p
\}_{p=1,\dots P}$ and $\{ \underline{\eta}^{(P)}_p \}_{p=1,\dots
P}$, identify two memoryless multi-mode bosonic channels
\cite{broadband}. Such channels are either attenuating channels, for
$\kappa \leqslant 1$, or amplifying channels for $\kappa > 1$. We
also remark that $\phi(\eta)\phi(\eta')=\phi(\eta\eta')$, that is,
the composition of two attenuating (amplifying) channels with
transmissivities (gains) $\eta$, $\eta'$, is an attenuating
(amplifying) channel with transmissivity (gain) $\eta''=\eta\eta'$.
Since the capacities do not increase under composition of channels,
we conclude that, for any $P$, the capacity of the memory channel is
bounded by the capacities of the memoryless multi-mode channels
identified by the set of parameters $\{ \overline{\eta}^{(P)}_p
\}_{p=1,\dots P}$ and $\{ \underline{\eta}^{(P)}_p \}_{p=1,\dots
P}$. (See Fig.\ \ref{bounds} for an illustrative example.)

As customary, to avoid unphysical results when discussing the
classical capacity of bosonic channels, we will adopt suitable
constraints on the input encodings (this can be avoided when dealing
with  the quantum capacity, since typically it does not diverge even
for unbounded inputs~\cite{HolevoWerner,Wolf}). Here we compute the
constrained classical capacity under the condition
\begin{equation}\label{constraint}
\frac{1}{n} \sum_{j=1}^n \Tr\left[ \rho^{(n)} a^\dag_j a_j \right]
\leqslant N \, ,
\end{equation}
where $\rho^{(n)}$ is the ensemble of input states over $n$ channel
uses, the parameter $N$ expresses the maximum number of mean
excitations per mode in average, and the constraint is intended to
hold for all values of $n$. It is important to notice that the
encoding transformation used for unraveling the memory channel
[Eq.~(\ref{a_collective})] preserves the input energy (this simply
follows from the fact that the matrix $O'$ is orthogonal). Thus, the
form of the energy constraint is preserved when written in terms of
the collective input variables $\{ \mathrm{a}_j, \mathrm{a}^\dag_j
\}$, i.e.,
\begin{equation}\label{constr_coll}
\frac{1}{n} \sum_{j=1}^n \Tr\left[ \rho^{(n)} \mathrm{a}^\dag_j
\mathrm{a}_j \right] \leqslant N \, .
\end{equation}

\subsection{Quantum capacity}\label{QCapacity}

The quantum capacity of the memoryless attenuating and amplifying
channels has been derived in \cite{Wolf}. By relying on additivity
arguments \cite{Caruso}, this result is readily extensible to the
case of memoryless multimode channels \cite{broadband}.

For the sake of simplicity, we focus on the limit of unbounded input
energy (i.e., $N\rightarrow \infty$), which leads to the
function~\cite{Wolf}
\begin{equation}\label{qfunction}
q(\eta) := \max \{ 0, \log_2{\eta}-\log_2{|\eta-1|} \} \, .
\end{equation}
We hence obtain, for any $P$, the expressions $\frac{1}{P} \sum_p
q[\overline{\eta}^{(P)}_p]$ and $ \frac{1}{P} \sum_p
q[\underline{\eta}^{(P)}_p]$ for the quantum capacity (per input
mode) of the multi-mode channels, respectively characterized by the
parameters $\overline{\eta}^{(P)}_p$ and $\underline{\eta}^{(P)}_p$.
Then, we can construct the following upper and lower bounds for the
quantum capacity of the memory channel:
\begin{subequations}
\begin{align}
\underline{Q}^{(P)} = & \left\{
\begin{array}{lr}
\frac{1}{P} \sum_p q[\underline{\eta}^{(P)}_p] \, , & \quad (\kappa \leqslant 1) \, , \\
\frac{1}{P} \sum_p q[\overline{\eta}^{(P)}_p] \, , & \quad (\kappa >
1) \, ,
\end{array}\right.\\
\overline{Q}^{(P)} = & \left\{
\begin{array}{lr}
\frac{1}{P} \sum_p q[\overline{\eta}^{(P)}_p] \, , & \quad (\kappa \leqslant 1) \, , \\
\frac{1}{P} \sum_p q[\underline{\eta}^{(P)}_p] \, , & \quad (\kappa
> 1) \, ,
\end{array}\right.
\end{align}
\end{subequations}
so that
\begin{equation}
\underline{Q}^{(P)} \leqslant Q \leqslant \overline{Q}^{(P)} \, .
\end{equation}
By varying the value of the integer $P$, we hence obtain a family of
lower and upper bounds on the quantum capacity $Q$ of the memory
channel.

In the limit $P\to\infty$, the collections of parameters $\{
\underline{\eta}^{(P)}_p\}$, $\{\overline{\eta}^{(P)}_p\}$ may
approach a limiting distribution. Clearly, this is the case when the
channel operates below threshold, with the limiting distribution
given in Eq.\ (\ref{mono_spectrum}). It follows that the upper and
lower bounds converge towards a same quantity in the limit
$P\to\infty$, and by applying the Szeg\"{o} theorem [Eq.\
(\ref{szego})] we can write
\begin{equation}\label{qcapacity}
Q = \int_0^{2\pi} \frac{dz}{2\pi} q[\eta(z)] \, .
\end{equation}
It is worth remarking that the Szeg\"{o} theorem can be applied on
smooth functions. Since the function in Eq.\ (\ref{qfunction}) is
singular for $\eta=1$, we deduce that the expression in Eq.\
(\ref{qcapacity}) coincides with the quantum capacity of the memory
channel if the parameters $\{ \underline{\eta}^{(P)}_p\}$,
$\{\overline{\eta}^{(P)}_p\}$ do not approach the unit value for all
the values of $P$ and $p$. By numerical evaluation of the
distribution of the parameters, we found evidences that this is the
case for the setting in which both initial and final memory modes
are assigned to the environment (see Sec.\ \ref{MUnraveling}). As
discussed in the Sec.\ \ref{Forget}, the expression in
(\ref{qcapacity}) does not coincide with the channel capacity if the
final memory mode is assigned to the receiver.

\begin{figure}[t]
\centering
\includegraphics[width=0.45\textwidth]{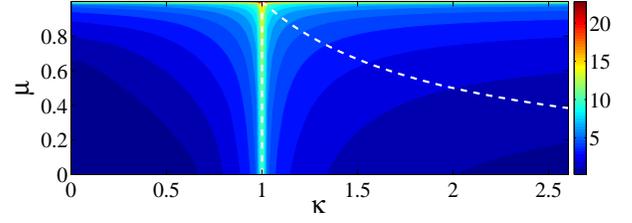}
\caption{(Color online.) The contour plot shows the quantum capacity
(measured in qubits per channel use) of the memory channel as
function of the amplifying/attenuating factor $\kappa$ and of the
memory parameter $\mu$. The quantum capacity is computed using Eq.\
(\ref{qcapacity}), which diverges logarithmically for $\mu \to 1$
and for $\kappa\to 1$. The dashed lines denote the boundaries among
different regions, from left to right: attenuating channel,
amplifying channel below threshold, amplifying channel above
threshold.} \label{Q}
\end{figure}

Above threshold, the distribution of eigenvalues is composed by a
continuous part, described by Eq.\ (\ref{mono_spectrum}), and by one
diverging eigenvalue. However, for the form of the function
(\ref{qfunction}), the latter does not contribute to the channel
capacity. In conclusion, the quantity in Eq.~(\ref{qcapacity})
expresses the quantum capacity of the memory channel both below and
above threshold, proving that $Q$ keeps no record of such
discontinuity. A plot of Eq.~(\ref{qcapacity}) as a function of  the
parameters $\mu$, $\kappa$ is reported in Fig.~\ref{Q} (for
$\kappa>1$, the dashed line represents the threshold value
$\kappa\mu=1$).

\subsection{Classical capacity}\label{CCapacity}

Similarly to the previous case, for the classical capacity we
consider a family of bounds constructed by introducing a proper set
of memoryless channels. For $P$ integer, this yields the
inequalities
\begin{equation}\label{IMPOC}
\underline{C}^{(P)} \leqslant C \leqslant \overline{C}^{(P)} \, ,
\end{equation}
where the bounds are computed from the classical capacity of the
memoryless multi-mode channels defined by the sequence of parameters
$\{ \underline{\eta}^{(P)}_p \}$, $\{ \overline{\eta}^{(P)}_p \}$.

Let us first consider the case of the attenuating memory channel
(i.e., $\kappa \leqslant 1$) for which exact results can be derived.
The constrained classical capacity of the memoryless multi-mode
channel has been derived in \cite{broadband}, from which we can
write
\begin{subequations}\label{Cbounds}
\begin{align}
\underline{C}^{(P)} & = \frac{1}{P} \sum_{p=1}^P g[\underline{\eta}^{(P)}_p \underline{N}_p] \, ,\\
\overline{C}^{(P)}  & = \frac{1}{P} \sum_{p=1}^P
g[\overline{\eta}^{(P)}_p \overline{N}_p] \, ,
\end{align}
\end{subequations}
where
\begin{equation}
g(x) := (x+1)\log_2{(x+1)} - x\log_2{x} \, .
\end{equation}
The positive parameters $\{ \underline{N}_p \}$, $\{ \overline{N}_p
\}$ describe the optimal distributions of the excitation numbers
over the collective input modes. For any $P$ the optimal
distributions can be computed by Lagrange method \cite{broadband}.
In terms of a Lagrange multiplier $L$, the optimal distribution is
\begin{equation}\label{Lagrange}
\underline{N}_p = \left\{ \underline{\eta}^{(P)}_p \left[
2^{L/\underline{\eta}^{(P)}_p} - 1 \right] \right\}^{-1} \, ,
\end{equation}
and the value of the multiplier is found accordingly to the
constraint (\ref{constr_coll}), which reads
\begin{equation}\label{Sum}
\frac{1}{P} \sum_{p=1}^P \underline{N}_p = N \, ,
\end{equation}
and analogously for the distribution $\{ \overline{N}_p \}$.

Taking the limit $P\to\infty$ and applying (\ref{szego}) we notice
that the two bounds (\ref{Cbounds}) converge to the same quantity.
Therefore we conclude that
\begin{equation}\label{classical}
C = \int_0^{2\pi} \frac{dz}{2\pi} g[\eta(z) N(z)] \, , \quad (\kappa
\leqslant 1) \, .
\end{equation}
The function $N(z)$ represents the optimal excitation number
distribution; by taking the limit of Eq.s (\ref{Lagrange}) and
(\ref{Sum}), it can be computed as
\begin{equation}
N(z) =\left\{\eta(z)\left[2^{L/\eta(z)}-1\right]\right\}^{-1},
\end{equation}
where the value of the Lagrange multiplier is determined by the
implicit equation
\begin{equation}
\int_0^{2\pi} \frac{dz}{2\pi} N(z) = N \, .
\end{equation}
In some limiting cases Eq.~(\ref{classical}) admits a close
analytical solution. For instance in the memoryless configuration
$\mu=0$, we get $\eta(z) = \kappa$, $N(z)=N$ and thus correctly
$C=g(\kappa N)$~\cite{broadband}. Vice-versa for $\kappa=1$
(noiseless channel) or $\mu=1$ (perfect memory channel) we have
$\eta(z)=1$, $N(z)=N$ and thus $C=g(N)$ (perfect transfer). Finally
for $\kappa=0$ (quantum shift channel) we get $\eta(z)=\mu$,
$N(z)=N$ and thus $C=g(\mu N)$. For generic values of the parameters
the resulting expression can be numerically evaluated, showing an
increase of $C$ for increasing memory $\mu$.

\begin{figure}[t]
\centering
\includegraphics[width=0.45\textwidth]{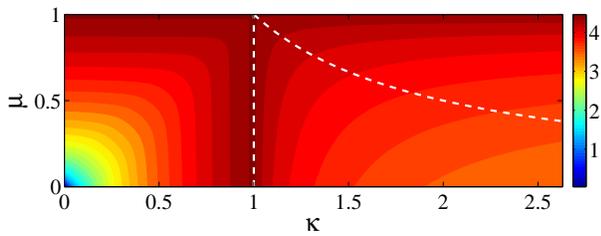}
\caption{(Color online.) The contour plot refers to the classical
capacity (measured in bits per channel use) of the memory channel as
function of the amplifying/attenuating factor $\kappa$ and of the
memory parameter $\mu$, under the constraint (\ref{constraint}) with
$N=8$. For the attenuation channel ($\kappa\leqslant 1$), the
classical capacity of the memory channel is plotted, using Eq.\
(\ref{classical}). For the amplifying channel ($\kappa > 1$), the
plotted quantity is a lower bound on the classical capacity. The
dashed lines denote the boundaries among different regions, from
left to right: attenuating channel, amplifying channel below
threshold, amplifying channel above threshold.} \label{C}
\end{figure}

Let us now consider the amplifying channel model, obtained for
$\kappa > 1$. This is intrinsically more complex than the previous
one, since in this case  no exact results are know even in the
memoryless case. Consequently Eq.~(\ref{IMPOC}) will only provide a
lower bound on the real capacity $C$ of our memory channel. For this
purpose we construct a lower bound for the classical capacity
$\underline{C}^{(P)}$ of the memoryless multi-mode channel by
restricting the coding strategy to only Gaussian
inputs~\cite{HolevoWerner} (it is worth noticing that such bound is
typically considered to be tight). This yields
\begin{equation}\label{lbound}
\underline{C}^{(P)} \geqslant  \frac{1}{P} \sum_{p=1}^P g\left[
\overline{\eta}^{(P)}_p(\overline{N}_p+1)+1\right] -
g\left[\overline{\eta}^{(P)}_p - 1\right] \, .
\end{equation}
The optimal distribution of the excitation numbers is computed by
Lagrange method. In terms of a Lagrange multiplier $L$, it reads
\begin{equation}
\overline{N}_p = \left\{ \overline{\eta}^{(P)}_p \left[
1-2^{-L/\overline{\eta}^{(P)}_p} \right] \right\}^{-1} - 1 \, .
\end{equation}
We have hence obtained a family of lower bounds on the memory
channel capacity. By taking the limit $P\to\infty$ we get
\begin{equation}\label{amp_low_bound}
C \geqslant \int_0^{2\pi} \frac{dz}{2\pi} \left( g\left\{
\eta(z)[N(z)+1]+1\right\} - g\left[\eta(z) - 1\right] \right) \, ,
\end{equation}
where $\eta(z)$ is the asymptotic function (\ref{mono_spectrum}),
and the optimal asymptotic distribution of excitation numbers is
given by
\begin{equation}
N(z) = \left\{ \eta(z) \left[ 1-2^{-L/\eta(z)} \right] \right\}^{-1}
- 1 \, .
\end{equation}
We stress that, as in the analysis of the quantum capacity $Q$, the
expression given in Eq.~(\ref{amp_low_bound}) holds also above
threshold. Indeed due to the form of the function~(\ref{lbound}),
the diverging eigenvalues give a vanishing contribution to the lower
bound on the classical capacity and can thus be neglected. Figure
\ref{C} shows the (lower bound on the) classical capacity of the
memory channel as function of the channel parameters $\mu$,
$\kappa$.

\section{Forgetfulness of the memory channel}\label{Forget}

According to Ref.\ \cite{KW2}, when dealing with memory channels one
has to distinguish among four different settings, which may lead to
different values of the channel capacities, depending whom the
initial and final memory modes are assigned to. We denote the four
possible settings by the label $XY$: $XY=AE$ (the initial memory
mode is assigned to the sender and the final one to the
environment), $AB$ (the initial and final memory modes are assigned
respectively to the sender and to the receiver), $EB$ (initial
memory mode to the environment and final one to the receiver), and
$EE$ (both initial and final memory modes are assigned to the
environment). If the initial memory mode is assigned to the
environment, the channel capacities may also depend on the memory
initialization. [We remind that the calculations presented in the
previous section have been performed in the  configuration $EE$
assuming the memory mode to be initialized in the vacuum].

The notion of {\it forgetfulness}, and hence the class of forgetful
memory channels, has been introduced in \cite{KW2} for quantum
channels acting on finite-dimensional Hilbert spaces. Forgetful
channels are characterized by several remarkable properties, in
particular the capacities of those channels do not depend on the
setting $XY$ and on the initialization of the memory mode. Let us
notice that our model is defined on an infinite-dimensional Hilbert
space, and hence the notion of forgetfulness cannot be directly
applied. However, we can argue on the forgetfulness of the channel
if an effective cutoff on the bosonic Hilbert space is introduced.
To prove the forgetfulness it is sufficient to show that, in the
limit of $n\to\infty$, the final state of the memory mode is
independent, in the sense specified in Ref.\ \cite{KW2}, on the
memory initialization. Indeed, the presence of the exponential
factor $\sqrt{\mu\kappa}^n$ in (\ref{m_lossy}), (\ref{m_ampl})
suggests that the channel is forgetful when $\mu\kappa < 1$, that
is, when operating below threshold. We can prove that this holds
true restricting to Gaussian states with bounded energy. To fix the
ideas, let us consider the case of the attenuating channel. The
transformation (\ref{m_lossy}) reads
\begin{equation}\label{m}
m'_n = \sqrt{\mu\kappa}^n \, m_1 + \mathbf{X}^\mathsf{T} \mathbf{a}
+ \mathbf{Y}^\mathsf{T} \mathbf{e} \, ,
\end{equation}
where $\mathbf{a}:=(a_1,\dots a_n)^\mathsf{T}$,
$\mathbf{e}:=(e_1,\dots e_n)^\mathsf{T}$, and the form of the
vectors $\mathbf{X}$, $\mathbf{Y}$ can be deduced from
(\ref{m_lossy}). A Gaussian state of the initial memory mode, the
input and the environmental modes, is characterized by the first
moments $\langle m_1 \rangle$, $\langle\mathbf{a}\rangle$,
$\langle\mathbf{e}\rangle$, and by the covariance matrix
\begin{eqnarray}
V = \left(\begin{array}{ccc}
V_m & C^\mathsf{T} & D^\mathsf{T} \\
C & V_\mathbf{a} & 0 \\
D & 0 & V_\mathbf{e}
\end{array}\right),
\end{eqnarray}
where the off-diagonal terms account for possible correlations of
the initial memory mode with the input and environment modes. After
$n$ uses of the channel, the state of the final memory mode is
Gaussian with first moment
\begin{equation}\label{1_moment}
\langle m'_n \rangle = \sqrt{\mu\kappa}^n \, \langle m_1 \rangle +
\mathbf{X}^\mathsf{T} \langle\mathbf{a}\rangle +
\mathbf{Y}^\mathsf{T} \langle\mathbf{e}\rangle \, ,
\end{equation}
and covariance matrix
\begin{align}\label{2_moment}
V_m' = (\mu\kappa)^n \, V_m & + \sqrt{\mu\kappa}^n \,
(\mathbf{X}^\mathsf{T}C + \mathbf{Y}^\mathsf{T}D +
C^\mathsf{T}\mathbf{X} + D^\mathsf{T}\mathbf{Y}) \nonumber\\
& + \mathbf{X}^\mathsf{T} V_\mathbf{a} \mathbf{X} +
\mathbf{Y}^\mathsf{T} V_\mathbf{e} \mathbf{Y} \, .
\end{align}
The presence of the exponential factors in (\ref{1_moment}),
(\ref{2_moment}) guarantees the forgetfulness of the channel as long
as the energy is bounded, e.g., subjected to a constraint of the
form
\begin{equation}
\frac{\langle m_1^\dag m_1 \rangle + \sum_{j=1}^n \langle a_j^\dag
a_j \rangle}{n+1} \le N.
\end{equation}
The same argument can be provided for the amplifying channel, as
long as it operates below threshold. It is finally worth reminding
that a first step towards the extension of the notion of
forgetfulness to the domain of continuous variables has been made in
\cite{weakforget}. However, we notice that this extension is not
straightforward if the full infinite dimensional Hilbert space is
considered. To illustrate this issue we consider the computation of
the quantum capacity, with unbounded input energy, in the generic
$XY$ setting. By inspection of Fig.s \ref{lossym}, \ref{amplifym},
it is immediate to see that in the $AB$ and $EB$ settings, where the
final memory mode is assigned to the receiver, at least one input
mode is perfectly transmitted, leading to an infinite value of the
quantum capacity for any value of the parameters $\mu$, $\kappa$. On
the other hand, we have shown in Sec.\ \ref{QCapacity} that the
quantum capacity is finite in the $EE$ setting (the same can be
shown in the $EB$ setting) for all values of the parameters,
provided $\mu, \kappa \neq 1$.

\section{Conclusions}\label{Conclusion}

We have introduced and characterized a model for memory effects in
attenuation and amplification quantum channels. Our findings show
that the presence of memory always increases the quantum capacity of
the communication line and may increase the classical one.

Interestingly enough, the highest rates of classical communication
are reached without the use of entangled codewords. For the
attenuating channel, the optimal encoding strategy for the
memoryless channels which bound $\Phi_n$ makes use of coherent
states~\cite{broadband}. Since the latter are preserved by the
encoding transformation $W_A$ our results prove, as a byproduct, the
optimality of coherent state encoding for the presented memory
channel.

We emphasis the use of the memory unraveling technique \cite{trans},
which may allows us to evaluate the channel capacity without relying
on the channel forgetfulness. This shows that the unraveling of the
memory could have a much broader impact than the results we have
presented here. In particular it may find applications on other
contexts, e.g., reaching beyond current restricted models that
involve statistically independent errors and are often inapplicable
to real physical systems. For instance, the model could be easily
adapted to describe situation in which temporal (causal)
correlations between the channel uses are replaced by spatial ones
or to deal with physical models where the underlying noise is more
complicated than attenuation (e.g., a stream of two level atoms
injected through a superconducting cavity). The model can also be
applied to describe memory effects in quantum repeaters and quantum
memories with atomic ensembles, where, via the Holstein-Primakoff
transformation \cite{H-P}, the interaction between matter and light
can be modeled as a formal beam-splitter (or linear-amplifier)
Hamiltonian \cite{interface}. An imperfect swap operation between
matter and light can be described by a beam-splitter with non-unit
transmissivity, which in turn may causes memory effects.

\appendix

\section{Explicit expressions}\label{APPE}

Writing the elements of the projector $P^{(n)}_{jj'}$ as $\psi_j
\psi_{j'}$, the explicit expressions for the quantities on the right
hand side of Eq.~(\ref{commuting}) are as follows
\begin{eqnarray}
c^{(n)} &:=&\tfrac{\kappa (\kappa -1) (1-\mu)^2}{(\mu \kappa -1)^2} \;\left[ 1 -\tfrac{1}{(\mu \kappa)^n} \right]  \nonumber \\
&& + \tfrac{\mu (\kappa -1)}{(\mu \kappa -1)^2} \; [ (\mu \kappa)^n -1]
- \tfrac{2 (1-\mu) (\kappa -1)}{\mu \kappa -1} \; n\;, \nonumber
\end{eqnarray}
\begin{eqnarray}
\psi_j := \tfrac{1}{\sqrt{c^{(n)}}} \; \left[ \sqrt{\tfrac{\kappa
-1}{\kappa (\mu \kappa -1)}} \sqrt{\mu \kappa}^j - (1-\mu) \;
\sqrt{\tfrac{\kappa (\kappa -1)}{\mu \kappa -1}} \tfrac{1}{\sqrt{\mu
\kappa}^j}\nonumber \right]  \;,
\end{eqnarray}
and
\begin{eqnarray}
\Delta M_{jj'}^{(n)} := \tfrac{(1-\mu)(\kappa-1)}{\mu \kappa -1} \tfrac{1}{\sqrt{\mu \kappa}^{|j-j'|}}
- (1-\mu)^2 \tfrac{\kappa (\kappa -1)}{\mu \kappa -1} \tfrac{1}{\sqrt{\mu \kappa}^{j+j'}} \;.\nonumber
\end{eqnarray}

\acknowledgments The research leading to these results has received
funding from the European Commission's seventh Framework Programme
(FP7/2007-2013) under grant agreements no.~213681, 
and by the Italian Ministry of University and Research under the
FIRB-IDEAS project RBID08B3FM.


\begin{thebibliography}{99}
\bibitem{NOTA0}
Similar results can be also derived for their entanglement assisted
counterparts.
\bibitem{network}
The methods developed in Sec.s \ref{MUnraveling}-\ref{Capacities}
may as well be applied to models of memory channels defined by
different combinations of the linear elements.
\bibitem{NOTA1}
The factor $1/2$ in the right hand side term of
Eq.~(\ref{mono_spectrum}) has been introduced to make $\eta(z)$
monotonic in $z$.
\end{thebibliography}
\end{document}